\documentclass[aps,onecolumn]{revtex4}
\usepackage{eurosym}
\usepackage{amsfonts}
\usepackage{amsmath}
\usepackage{amsopn}
\usepackage{amssymb}
\usepackage{graphicx}
\usepackage{graphics}
\usepackage{color}
\usepackage{verbatim}

\begin{document}

\title{Dynamics of dipoles and vortices in nonlinearly-coupled
three-dimensional field oscillators}
\author{R. Driben$^{1}$, V. V. Konotop$^{2}$, B. A. Malomed$^{3}$, and T.
Meier$^{1}$}
\affiliation{$^{1}$Department of Physics and CeOPP, University of Paderborn, Warburger
Str. 100, D-33098 Paderborn, Germany\\
$^{2}$Centro de F\'isica Te\'orica e Computacional and Departamento de
F\'isica, Faculdade de Ci\^encias, Universidade de Lisboa, Campo Grande,
Edif\'icio C8, Lisboa 1749-016, Portugal\\
$^{3}$Department of Physical Electronics, School of Electrical Engineering,
Faculty of Engineering, Tel Aviv University, Tel Aviv 69978, Israel}

\begin{abstract}
The dynamics of a pair of harmonic oscillators (HOs) represented by
three-dimensional fields coupled by a repulsive cubic nonlinearity is
investigated through direct simulations of the respective field equations,
and with the help of the finite-mode Galerkin approximation (GA), which
represents the two interacting fields by a superposition of $3+3$ HO \textit{%
p}-wave eigenfunctions with orbital and magnetic quantum numbers $l=1$ and $%
m=1,0,-1$.
The system can be implemented in binary Bose-Einstein condensates,
demonstrating a potential of the atomic condensates for emulating various
complex modes predicted by classical field theories, First, the GA very
accurately predicts a \textit{broadly degenerate} set of the system's ground
states in the \textit{p}-wave manifold, in the form of complexes built of a
dipole coaxial with another dipole or vortex, as well as complexes built of
mutually orthogonal dipoles. Next, pairs of non-coaxial vortices and/or
dipoles, including pairs of mutually perpendicular vortices, develop
remarkably stable dynamical regimes, which feature periodic exchange of the
angular momentum and periodic switching between dipoles and vortices. For a
moderately strong nonlinearity, simulations of the coupled field equations
agree very well with results produced by the GA, demonstrating that the
dynamics is accurately spanned by the set of six modes limited to $l=1$.
\end{abstract}

\maketitle


\section{Introduction}

The study of diverse complex three-dimensional (3D) modes, such as spinning
solitons and vortex rings \cite{vortex-rings}, knots \cite{knots}, hopfions
\cite{hopfions}, and skyrmions \cite{skyrmions} is one of central topics in
the classical field theory \cite{review}. In addition to the well-known
applications of skyrmions to low-energy hadron physics \cite{hadrons},
states approximated by the field-theory modes can be identified in
ferromagnets and ferroelectrics \cite{ferro}, superconductors \cite{super},
and semiconductors \cite{semi}.

An exceptionally clean and well-controlled implementation of various modes
predicted in the field theory is offered by atomic Bose-Einstein condensates
(BECs). In particular, the possibility of the creation of BEC\ skyrmions was
predicted in a number of settings \cite{BEC-skyrmions} and realized
experimentally \cite{skyrmion-exper}. Also predicted were settings
appropriate for the creation of atomic knots \cite{BEC-knots} and hopfions
\cite{BEC-hopfion}, and the experimental creation of monopoles was recently
reported in Ref. \cite{monopole-exper}.

This topic is closely related to the general problem of the creation of 2D
and 3D self-trapped modes in nonlinear media (often considered as solitons),
which also finds many physically relevant ramifications \cite%
{general-reviews}, \cite{RMP}, especially in nonlinear optics and
matter-wave dynamics in BECs. In addition to the great significance to
fundamental studies in these fields, 2D optical solitons may be used as bit
carriers in all-optical data-processing schemes \cite{Wagner}, and 3D
matter-wave solitons are expected to provide a basis for precise
interferometry \cite{interferometry}.

The objective of this work is to demonstrate that a 3D two-component system,
which can be readily implemented in BEC, makes it possible to create new
nontrivial bound states and robust dynamical regimes, built of coaxial or
mutually perpendicular dipole modes or vortices in the two components, that
are of definite interest as field-theory modes, and, simultaneously, suggest
novel possibilities for the experimental emulation of the field-theory
settings in atomic BEC. Although the overall 3D nonlinear-field dynamics
seems complex enough (remaining regular, rather then getting chaotic), a
remarkable fact is that it can be very accurately represented by a six-mode
Galerkin approximation (GA), which, in turn, admits an essentially
analytical investigation. This finding suggests that the projection onto an
appropriately chosen GA may be useful in other nonlinear field-theory models
too.

Getting back to the introduction into the general topic, it is relevant to
mention that the creation of multidimensional solitons is often complicated
by the fact that such states, usually supported by the cubic self-focusing
nonlinearity, are subject to instability caused by the wave collapse
(critical and supercritical collapse in 2D and 3D, respectively) \cite%
{collapse}. Multidimensional vortex solitons, alias vortex rings, in cubic
media are vulnerable to a still stronger splitting instability induced by
modulational perturbations in the azimuthal direction \cite{general-reviews}%
. These instabilities may be suppressed by means of periodic potentials
(lattices), as was predicted theoretically \cite{lattice} and demonstrated
in the experiment \cite{photorefr}; very recently, it was also predicted
that the spin-orbit coupling can stabilize 2D \cite{Fukuoka} and 3D \cite{HP}
solitons in two-component systems.

Such physical media are modeled by the non-relativistic nonlinear Schr\"{o}%
dinger equation, alias Gross-Pitaevskii equation (GPE) \cite{Pit}, for the
corresponding photonic or atomic mean-field wave function $\Phi \left(
\mathbf{r},t\right) $, where $\mathbf{r}=(x,y,z)$ and $t$ are appropriately
scaled spatial coordinates and time:%
\begin{equation}
i\frac{\partial \Phi }{\partial t}=-\nabla ^{2}\Phi +U\left( \mathbf{r}%
\right) \Phi +\sigma |\Phi |^{2}\Phi .  \label{GPE}
\end{equation}%
In this equation, $\nabla ^{2}$ is the 3D Laplacian, $U\left( \mathbf{r}%
\right) $ is the trapping potential, and $\sigma =-1$ or $+1$ corresponds to
the self-focusing or defocusing (repulsive) sign of the cubic interaction.\
In particular, in the case of axially symmetric potentials (including
spherically isotropic ones), $U=U\left( \rho ,z\right) $, where $\left( \rho
,z,\varphi \right) $ is the set of cylindrical coordinates, Eq.~(\ref{GPE})
admits vortical modes in the form of%
\begin{equation}
\Phi =\exp \left( -i\mu t+il\varphi \right) u(\rho ,z),  \label{vort}
\end{equation}%
with real chemical potential $\mu $, integer vorticity $l$, and real
amplitude function $u\left( \rho ,z\right) $ satisfying the stationary
equation,%
\begin{equation}
\mu u=-\left( \frac{\partial ^{2}}{\partial \rho ^{2}}+\frac{1}{\rho }\frac{%
\partial }{\partial \rho }+\frac{\partial ^{2}}{\partial z^{2}}-\frac{l^{2}}{%
\rho ^{2}}\right) u+U\left( \rho ,z\right) u+\sigma u^{3}.  \label{u}
\end{equation}

A novel approach for the creation of self-trapped modes was proposed in Ref.
\cite{ICFO}: A \textit{spatially inhomogeneous} repulsive cubic
nonlinearity, whose local strength in the space of dimension $D$ grows from
the center to periphery faster than $r^{D}$ ($r$ is the radial coordinate),
supports extremely robust and diverse families of solitons, multipoles,
fundamental and composite solitary vortices, and hopfions for $D=1,2,3$ \cite%
{further}. This type of the nonlinearity modulation belongs to the general
class of the nonlinear pseudopotentials, which can be induced by various
techniques in optics for $D=1,2$, and in BEC for $D=3$ too \cite{RMP}.

On the other hand, it was demonstrated in detail theoretically that the
usual 2D and 3D settings, which combine spatially uniform cubic
self-repulsion and an isotropic harmonic-oscillator (HO) trapping potential,
readily give rise to various bound states, including trapped vortices \cite%
{vortices} as well as vortex clusters and dipoles \cite{vort-clusters}. The
formation of vortices in optics and BEC\ was reported in many experimental
works \cite{vort-exper}, see also review \cite{vort-rev}. The stability of
the vortices in these settings is essentially secured by the fact that the
repulsive nonlinearity does not give rise to modulational instability.

Multi-component BECs with repulsive self- and cross-interactions (usually,
they are realized as mixtures of different hyperfine atomic states of the
same species) may also give rise to stable vortices \cite{2comp-vort} and,
furthermore, to the above-mentioned matter-wave skyrmions \cite%
{BEC-skyrmions,skyrmion-exper} and monopoles \cite{monopole-exper}. Further,
the studies of two-component systems suggest that two identical 3D\ GPEs
coupled by the repulsive interaction, each with an isotropic HO trapping
potential, may be used as a simple model for the analysis of nontrivial
dynamical regimes, such as interaction of non-coaxial (or, more
specifically, mutually perpendicular) trapped vortices or dipoles initially
created in the two wave fields. This possibility was recently proposed in
Ref.~\cite{arxiv} were stable vortices having orthogonal vortex lines and
trapped in a HO trap were found. If initially the modes deviate from the
stationary state, the nonlinear repulsive interactions lead to smooth
dynamics representing torque-free precession with nutations. The model was
studied under the condition that the inter-species repulsion was weaker than
the self-repulsion in each component. In this case, it was concluded that
the system is robust with respect to variation of parameters.

In the present work we address a system of two GPEs which is dominated by
intra-species repulsion, while the intra-species self-interactions may be
neglected. We predict experimentally observable stationary and dynamical
modes by means of systematic simulations and, in parallel, making use of the
above-mentioned six-mode GA. The approach developed here is similar to its
well-known applications of the GA in hydrodynamics \cite{GA}, projecting the
coupled GPEs for two mean-field wave functions onto the truncated dynamical
system. The six-mode truncation approximates each wave function by a
combination of three \textit{p}-wave eigenfunctions of the 3D HO, with
orbital quantum number $l=1$ and magnetic quantum numbers $m=-1,0,+1$,
assuming that the axes which define the two triplets of the eigenfunctions
are \emph{mutually perpendicular}. The accuracy provided by the GA turns out
to be surprisingly high, provided that the nonlinear interaction is not too
strong (at certain threshold of nonlinear interaction strength a deviation from the GA is observed
due to generation of components with $l>1$). In particular, fixed points
(FPs) of the GA provide for a very good approximation for quasi-stationary
states of the coupled GPE system. Taking sets of non-coaxial dipoles or
vortices in the two wave fields as initial states, their nonlinear
interaction leads, in the framework of the GPEs and GA alike, to remarkably
robust dynamical regimes, which are characterized by a periodic exchange of
the angular momentum between the two field components and a periodic switch
of their structure, with the dipoles transforming into vortices and vice
versa.

The model and GA are introduced in Sec.~\ref{model}, which is followed by
the analysis of the GA's FPs in Sec. \ref{FP}. In the same section, we also
produce (energy-degenerate) ground states (GSs) of the coupled GPEs in the
\textit{p}-wave manifold, which are very accurately predicted by the FPs of
the GA. In Sec.~\ref{numres} we present systematic results for the dynamics
of pairs of non-coaxial nonlinearly coupled dipoles and/or vortices. The
paper is concluded by Sec.~\ref{sum}.

\section{The model and the Galerkin approximation}

\label{model}

The scaled form of the underlying system of the coupled 3D GPEs with the
repulsive interaction between the two wave functions, $\Phi $ and $\Psi $,
and the isotropic HO potential, represented by terms $\sim r^{2}$, is%
\begin{eqnarray}
i\frac{\partial \Phi }{\partial t} &=&-\nabla ^{2}\Phi +r^{2}\Phi +|\Psi
|^{2}\Phi ,  \label{phi} \\
i\frac{\partial \Psi }{\partial t} &=&-\nabla ^{2}\Psi +r^{2}\Psi +|\Phi
|^{2}\Psi .  \label{psi}
\end{eqnarray}%
Unlike Ref. \cite{arxiv}, we here consider the system with negligible
self-repulsion of each component in comparison with the repulsive
interaction between them. In the BEC\ experiment, this situation may be
achieved using the Feshbach resonance (FR)\ induced by external magnetic
field to strengthen the inter-component repulsion \cite%
{Feshbach+rf,Feshbach2}, thus making this interaction much stronger than the
intra-component interactions. The effect of the FR may be additionally
enhanced if applied to atomic states \textquotedblleft dressed" by a
radio-frequency field \cite{Feshbach+rf,Shlyap}). In particular, it was
demonstrated experimentally \cite{Feshbach2} that the FR can make the
scattering length accounting for the repulsion between atomic states of $%
^{87}$Rb with quantum numbers $\left\vert F=1,m_{F}=+1\right\rangle $ and $%
\left\vert F=2,m_{F}=-1\right\rangle $, at magnetic field $9.10$ G, larger
by a factor $\simeq 15$ than the scattering length which represents the
intra-component self-repulsion, cf. Ref. \cite{Australia}. In fact,
simulations of equations obtained from Eqs. (\ref{phi}), (\ref{psi}) by
adding self-repulsion terms, i.e., $\epsilon |\Phi |^{2}\Phi $ and $\epsilon
|\Psi |^{2}\Psi $, respectively produce results(not shown here in detail) which are qualitatively similar to those presented in the current
work for $\epsilon =0$. Only quantitative characteristics will be fifferent such as duration of nutations as well as the duration of the total round trip. Pure numerical examples with SPM included for a pair of vortices are presented in \cite{arxiv}.

Equations (\ref{phi}) and (\ref{psi}) conserve two norms,
\begin{equation}
N_{\Phi }=\int \left\vert \Phi (\mathbf{r}))\right\vert ^{2}d\mathbf{r}%
,\qquad N_{\Psi }=\int \left\vert \Psi (\mathbf{r})\right\vert ^{2}d\mathbf{r%
},  \label{N-N}
\end{equation}%
%
%
%
%
%
%
%
the total vectorial angular momentum, $\mathbf{M}=\mathbf{M}_{\Phi }+\mathbf{%
M}_{\Psi }$ where
\begin{equation}
\mathbf{M}_{\Phi }=-i\int \Phi ^{\ast }\left( \mathbf{r}\times \nabla
\right) \Phi \,d\mathbf{r},\qquad \mathbf{M}_{\Psi }=-i\int \Psi ^{\ast
}\left( \mathbf{r}\times \nabla \right) \Psi \,d\mathbf{r},  \label{M}
\end{equation}%
and the Hamiltonian,%
\begin{equation}
H=\int \left[ \left\vert \nabla \Phi \right\vert ^{2}+\left\vert \nabla \Psi
\right\vert ^{2}+r^{2}\left( |\Phi |^{2}+|\Psi |^{2}\right) \right] \,d%
\mathbf{r}+E_{\mathrm{int}},  \label{H}
\end{equation}%
which includes the interaction energy,%
\begin{equation}
E_{\mathrm{int}}=\int \left\vert \Phi (\mathbf{r})\right\vert ^{2}\left\vert
\Psi (\mathbf{r})\right\vert ^{2}\,d\mathbf{r}.  \label{Eint}
\end{equation}%
%
%
%
%
%
%
%
Note that, for stationary vortical modes, with%
\begin{equation}
\left\{ \Phi ,\Psi \right\} =\exp \left( -i\mu t+il\varphi \right) \left\{
u(\rho ,z),v\left( \rho ,z\right) \right\}  \label{vortvort}
\end{equation}
(cf. Eq. (\ref{vort})), it follows from Eq.~(\ref{M}) that the absolute
value of the total angular momentum is a multiple of the norm,
\begin{equation}
M_{\Phi,\Psi}=lN_{\Phi,\Psi},  \label{l}
\end{equation}%
irrespective of the particular structure of modal functions $u\left( \rho
,z\right) $ and $v\left( \rho ,z\right) $ in Eq. (\ref{vortvort}).

As said above, the GA represents the two wave functions, $\Phi $ and $\Psi $%
, as superpositions of three \textit{p}-wave HO eigenstates, $\left( 1/\sqrt{%
C_{lm}}\right) re^{-r^{2}/2}Y_{l}^{m}\left( \mathbf{r}\right) $, where $%
Y_{l}^{m}$ are spherical harmonics (written in terms of the Cartesian
coordinates; recall that $r=\sqrt{x^{2}+y^{2}+z^{2}}$) with quantum numbers $%
l=1$, $m=1,0,-1$, and normalization constants\ $C_{lm}=\int \mathbf{dr~}%
r^{2}e^{-r^{2}}\left\vert Y_{l}^{m}\left( \mathbf{r}\right) \right\vert ^{2}$%
. We define the GA so that the vorticity axes for the two triplets of the
eigenfunctions are aligned with perpendicular coordinate directions, $z$ and
$y$ (cf. Ref.~\cite{arxiv}). Thus we use the following \textit{Ans\"{a}tze},
in which the normalized HO eigenfunctions are substituted in their explicit
forms, and $a_{1,0,-1}(t)$ and $b_{1,0,-1}(t)$ are the expansion amplitudes:
\begin{eqnarray}
\Phi \left( \mathbf{r},t\right) &=&\frac{1}{\pi ^{3/4}}e^{-5it}re^{-r^{2}/2}%
\left[ a_{1}(t)\frac{x+iy}{r}+\sqrt{2}a_{0}(t)\frac{z}{r}+a_{-1}(t)\frac{x-iy%
}{r}\right] ,  \label{phiGA} \\
\Psi \left( \mathbf{r},t\right) &=&\frac{1}{\pi ^{3/4}}e^{-5it}re^{-r^{2}/2}%
\left[ b_{1}(t)\frac{x-iz}{r}+\sqrt{2}b_{0}(t)\frac{y}{r}+b_{-1}(t)\frac{x+iz%
}{r}\right] .  \label{psiGA}
\end{eqnarray}%
Then, the evolution equations for the amplitudes are produced by the
projection of the GPEs (\ref{phi}) and (\ref{psi}) onto the set of the
eigenfunctions. After some algebra, the following dynamical system with six
degrees of freedom is thus derived [$\tau \equiv t/\left( 16\sqrt{2}\pi
^{3/2}\right) $]:%
\begin{gather}
i\frac{d{a}_{1}}{d\tau }%
=2|b_{0}|^{2}(2a_{1}-a_{-1})+(|b_{1}|^{2}+|b_{-1}|^{2})(3a_{1}+a_{-1})+
\notag \\
+a_{0}\left[ b_{0}(b_{1}^{\ast }-b_{-1}^{\ast })+b_{0}^{\ast }(b_{-1}-b_{1})%
\right] +(a_{1}+a_{-1})(b_{1}b_{-1}^{\ast }+b_{1}^{\ast }b_{-1})-  \notag \\
-i\sqrt{2}\left[ a_{0}(b_{-1}^{\ast }b_{1}-b_{-1}b_{1}^{\ast
})+a_{-1}(b_{0}b_{-1}^{\ast }+b_{0}b_{1}^{\ast }+b_{0}^{\ast
}b_{-1}+b_{0}^{\ast }b_{1})\right] ,  \notag \\
i\frac{d{a}_{0}}{d\tau }%
=2a_{0}(|b_{0}|^{2}+2|b_{-1}|^{2}+2|b_{1}|^{2})-2a_{0}\left(
b_{1}b_{-1}^{\ast }+b_{1}^{\ast }b_{-1}\right) -  \notag \\
-(a_{1}-a_{-1})(b_{0}(b_{1}^{\ast }-b_{-1}^{\ast })-b_{0}^{\ast
}(b_{1}-b_{-1}))-\sqrt{2}i(a_{-1}+a_{1})\left( b_{1}b_{-1}^{\ast
}-b_{1}^{\ast }b_{-1}\right) ,  \notag \\
i\frac{d{a}_{-1}}{d\tau }%
=2|b_{0}|^{2}(2a_{-1}-a_{1})+(|b_{1}|^{2}+|b_{-1}|^{2})(3a_{-1}+a_{1})-
\notag \\
-a_{0}\left[ b_{0}(b_{1}^{\ast }-b_{-1}^{\ast })+b_{0}^{\ast }(b_{-1}-b_{1})%
\right] +(a_{1}+a_{-1})(b_{1}b_{-1}^{\ast }+b_{1}^{\ast }b_{-1})+  \notag \\
+i\sqrt{2}\left[ a_{0}(b_{-1}b_{1}^{\ast }-b_{-1}^{\ast
}b_{1})+a_{1}(b_{0}b_{-1}^{\ast }+b_{0}b_{1}^{\ast }+b_{0}^{\ast
}b_{-1}+b_{0}^{\ast }b_{1})\right] ;  \label{a}
\end{gather}

\begin{gather}
i\frac{d{b}_{1}}{d\tau }%
=2|a_{0}|^{2}(2b_{1}-b_{-1})+(|a_{1}|^{2}+|a_{-1}|^{2})(3b_{1}+b_{-1})+
\notag \\
+b_{0}\left[ a_{0}(a_{1}^{\ast }-a_{-1}^{\ast })+a_{0}^{\ast }(a_{-1}-a_{1})%
\right] +(b_{1}+b_{-1})(a_{1}a_{-1}^{\ast }+a_{1}^{\ast }a_{-1})+  \notag \\
+i\sqrt{2}\left[ b_{0}(a_{-1}^{\ast }a_{1}-a_{-1}a_{1}^{\ast
})+b_{-1}(a_{0}a_{-1}^{\ast }+a_{0}a_{1}^{\ast }+a_{0}^{\ast
}a_{-1}+a_{0}^{\ast }a_{1})\right] ,  \notag \\
i\frac{d{b}_{0}}{d\tau }%
=2b_{0}(|a_{0}|^{2}+2|a_{-1}|^{2}+2|a_{1}|^{2})-2b_{0}\left(
a_{1}a_{-1}^{\ast }+a_{1}^{\ast }a_{-1}\right) -  \notag \\
-(b_{1}-b_{-1})(a_{0}(a_{1}^{\ast }-a_{-1}^{\ast })-a_{0}^{\ast
}(a_{1}-a_{-1}))+\sqrt{2}i(b_{-1}+b_{1})\left( a_{1}a_{-1}^{\ast
}-a_{1}^{\ast }a_{-1}\right) ,  \notag \\
i\frac{d{b}_{-1}}{d\tau }%
=2|a_{0}|^{2}(2b_{-1}-b_{1})+(|a_{1}|^{2}+|a_{-1}|^{2})(3b_{-1}+b_{1})-
\notag \\
-b_{0}\left[ a_{0}(a_{1}^{\ast }-a_{-1}^{\ast })+a_{0}^{\ast }(a_{-1}-a_{1})%
\right] +(b_{1}+b_{-1})(a_{1}a_{-1}^{\ast }+a_{1}^{\ast }a_{-1})-  \notag \\
-i\sqrt{2}\left[ b_{0}(a_{1}^{\ast }a_{-1}-a_{-1}^{\ast
}a_{1})+b_{1}(a_{0}a_{-1}^{\ast }+a_{0}a_{1}^{\ast }+a_{0}^{\ast
}a_{-1}+a_{0}^{\ast }a_{1})\right] .  \label{b}
\end{gather}

It is straightforward to check that Eqs. (\ref{a}) and (\ref{b}) conserve
the two norms, which exactly correspond to expressions (\ref{N-N}),%
\begin{equation}
N_{a}=|a_{-1}|^{2}+|a_{0}|^{2}+|a_{1}|^{2},\quad
N_{b}=|b_{-1}|^{2}+|b_{0}|^{2}+|b_{1}|^{2},  \label{NaNb}
\end{equation}%
and the equations can be written in the canonical Hamiltonian form%
\begin{equation}
i\frac{da_{j}}{d\tau }=\frac{\partial H}{\partial a_{j}^{\ast }},\quad i%
\frac{db_{j}}{d\tau }=\frac{\partial H}{\partial b_{j}^{\ast }},  \label{HH}
\end{equation}%
$j=1,0,-1$, where the conserved Hamiltonian can be cast in the following
form, taking into account the definition of the norms (\ref{NaNb}):
\begin{gather}
H=3N_{a}N_{b}+|b_{0}|^{2}N_{a}+|a_{0}|^{2}N_{b}+  \notag \\
+(|a_{-1}|^{2}+|a_{1}|^{2}-2|a_{0}|^{2})(b_{1}b_{-1}^{\ast }+b_{1}^{\ast
}b_{-1})+(|b_{-1}|^{2}+|b_{1}|^{2}-2|b_{0}|^{2})(a_{1}a_{-1}^{\ast
}+a_{1}^{\ast }a_{-1})-  \notag \\
+\left[ a_{0}^{\ast }(a_{-1}-a_{1})+a_{0}(a_{1}^{\ast }-a_{-1}^{\ast })%
\right] \left[ b_{0}(b_{1}^{\ast }-b_{-1}^{\ast })+b_{0}^{\ast
}(b_{-1}-b_{1})\right] +  \notag \\
+(a_{1}a_{-1}^{\ast }+a_{-1}a_{1}^{\ast })(b_{1}b_{-1}^{\ast
}+b_{-1}b_{1}^{\ast })-  \notag \\
-i\sqrt{2}\left[ a_{0}(a_{1}^{\ast }+a_{-1}^{\ast })(b_{-1}^{\ast
}b_{1}-b_{-1}b_{1}^{\ast })+b_{0}(b_{1}^{\ast }+b_{-1}^{\ast
})(a_{-1}a_{1}^{\ast }-a_{-1}^{\ast }a_{1})\right. +  \notag \\
\left. +a_{0}^{\ast }(a_{1}+a_{-1})(b_{-1}b_{1}^{\ast }-b_{-1}^{\ast
}b_{1})+b_{0}^{\ast }(b_{1}+b_{-1})(a_{-1}a_{1}^{\ast }-a_{-1}^{\ast }a_{1})
\right] .  \label{HHH}
\end{gather}%
This Hamiltonian corresponds to the substitution of the \textit{Ans\"{a}tze}
(\ref{phiGA}), (\ref{psiGA}) into the interaction energy (\ref{Eint}) (the
non-interaction terms in the Hamiltonian (\ref{H}) are removed from its GA
counterpart (\ref{HHH}) through the definition of the GA \textit{Ans\"{a}tze}%
, which eliminates the dynamics determined by those terms).

\section{Fixed points of the Galerkin approximation and ground states of the
coupled Gross-Pitaevskii equations for $l=1$}

\label{FP}

\begin{figure}[tbh]
\centering
\includegraphics[width=9cm]{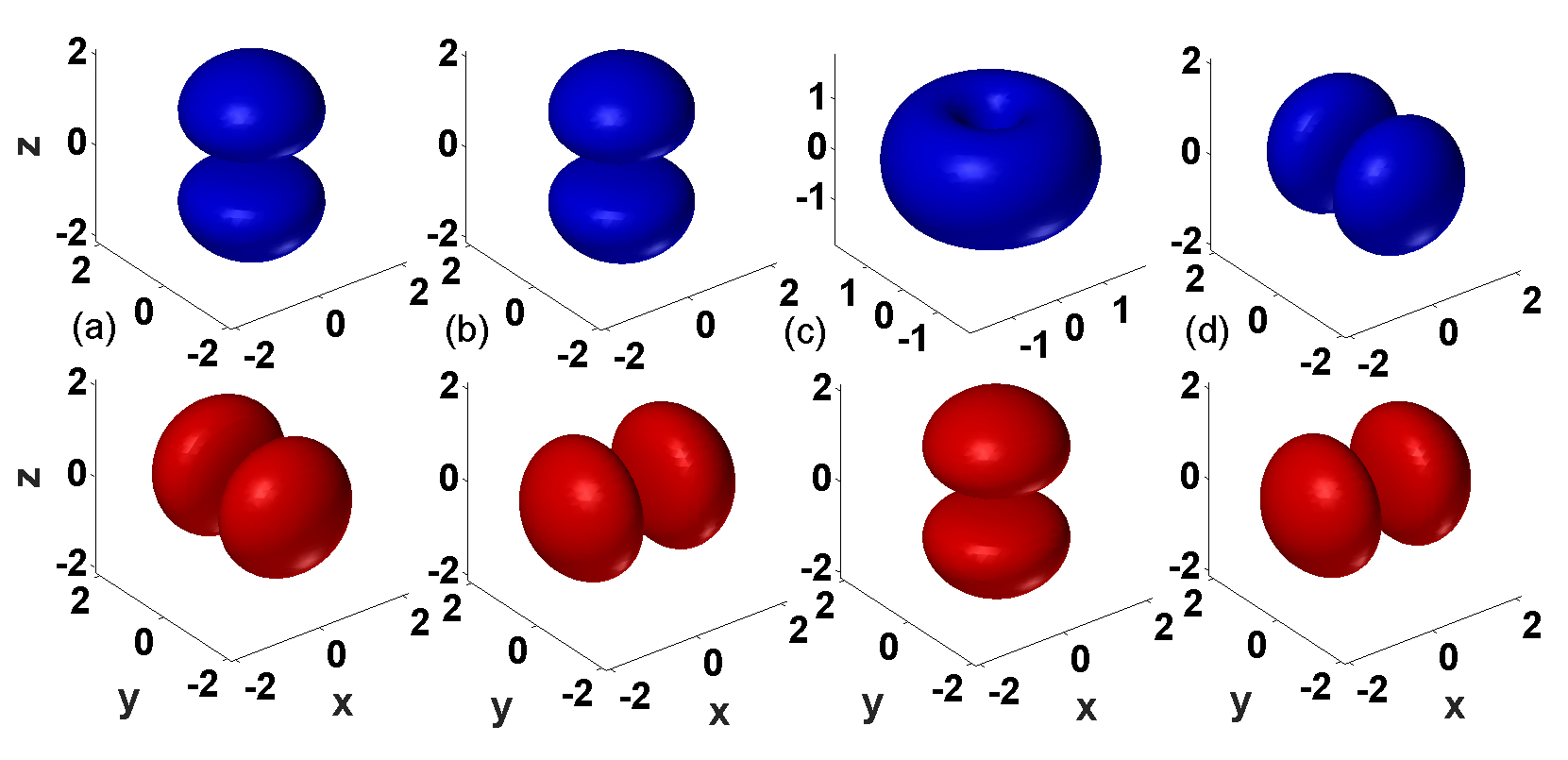}
\caption{(Color online) Density isosurfaces, with $|\Phi |^{2}=|\Psi |^{2}=%
\mathrm{const}$, represent shapes of the two components (the top and bottom
rows, respectively) of stationary modes that correspond to fixed points (%
\protect\ref{DD}), (\protect\ref{ODD}), (\protect\ref{VD}), and (\protect\ref%
{aab}) of the Galerkin approximation, in panels (a), (b), (c), and (d),
respectively. These are orthogonal dipoles with different orientations in
(a), (b) and (d), and the coaxial vortex-dipole complex in (c).}
\label{fig1}
\end{figure}

It is easy to find FPs of Eqs. (\ref{a}) and (\ref{b}) as the following
stationary solutions. The first solution is

\begin{equation}
\left( a_{1},a_{0},a_{-1},b_{1},b_{0},b_{-1}\right) =\left( 0,a,0,\frac{a}{%
\sqrt{2}},0,\frac{a}{\sqrt{2}}\right) e^{-2ia^{2}t}.  \label{DD}
\end{equation}%
%
%
%
%
%
%
%
\textit{Ans\"{a}tze} (\ref{phiGA}), (\ref{psiGA}) demonstrate, as plotted in
the left column of Fig. \ref{fig1}, that this FP corresponds, in terms of
the full wave functions, to a coaxial dipole-dipole (CDD) mode, although the
respective dipoles in the $\Phi $ and $\Psi $ wave functions are of
different types: the former one is represented by the HO eigenfunction, with
$l=1,m=0$, defined as a \textit{p}-wave with respect to vorticity axis $z$,
while the latter one is a combination of two \textit{p}-waves with $%
l=1,m=\pm 1$, defined with respect to axis $y$. The substitution of (\ref{DD}%
) into (\ref{phiGA}) and (\ref{psiGA}) yields:
\begin{equation*}
\Phi \left( \mathbf{r},t\right) =\frac{\sqrt{2}a}{\pi ^{3/4}}%
e^{-5it}ze^{-r^{2}/2},~\Psi \left( \mathbf{r},t\right) =\frac{\sqrt{2}a}{\pi
^{3/4}}e^{-5it}xe^{-r^{2}/2}.
\end{equation*}%
These expression clearly represent orthogonal dipoles, with one oriented
along $z$-axis and the other one oriented along the $x$-axis as shown in the
left column of Fig. \ref{fig1}(a). The value of Hamiltonian (\ref{HHH}) for
this FP is%
\begin{equation}
H_{\mathrm{CDD}}=2a^{4}.  \label{HCD}
\end{equation}

The second FP is
\begin{equation}
\left( a_{1},a_{0},a_{-1},b_{1},b_{0},b_{-1}\right) =\left(
0,a,0,0,a,0\right) e^{-2ia^{2}t}.  \label{ODD}
\end{equation}%
%
%
%
%
%
%
%
According to the \textit{Ans\"{a}tze} (\ref{phiGA}), (\ref{psiGA}), this FP
corresponds to a mode displayed in Fig. \ref{fig1}(b). It is composed of
mutually orthogonal dipoles (ODD), each one being represented by the HO
\textit{p}-wave with $l=1,m=0$, defined with respect to the vorticity axes $%
z $ and $y$. The Hamiltonian of this FP is%
\begin{equation}
H_{\mathrm{ODD}}=5a^{4}.  \label{HOD}
\end{equation}

The third FP is
\begin{equation}
\left( a_{1},a_{0},a_{-1},b_{1},b_{0},b_{-1}\right) =\left( a,0,0,\frac{a}{%
\sqrt{2}},0,-\frac{a}{\sqrt{2}}\right) e^{-2ia^{2} t},  \label{VD}
\end{equation}%
%
%
%
%
%
%
%
Through \textit{Ans\"{a}tze} (\ref{phiGA}), (\ref{psiGA}), it corresponds to
a coaxial combination of the vortex in $\Phi $ (the single nonzero amplitude
$a_{1}$ clearly represents a vortex with respect to axis $z$) and dipole in $%
\Psi $, as shown in Fig. \ref{fig1}(c). The Hamiltonian of this coaxial
vortex-dipole (CVD) FP is%
\begin{equation}
H_{\mathrm{CVD}}=2a^{4}.  \label{HVD}
\end{equation}

The calculation of eigenfrequencies of small perturbations around all these
FPs, in the framework of the linearized version of Eqs. (\ref{a}) and (\ref%
{b}) (i.e., the respective Bogoliubov - de Gennes equations \cite{Pit}),
demonstrates that all the FPs are stable, as all eigenfrequencies are real.
Furthermore, the norms (\ref{NaNb}) of all the three FPs given by Eqs. (\ref%
{DD}) (\ref{ODD}) and (\ref{VD}), with fixed $a$, are equal: $%
N_{a}=N_{b}=a^{2}$. Therefore, it makes sense to compare the respective
values of the Hamiltonian, given by Eqs. (\ref{HCD}), (\ref{HOD}) and (\ref%
{HVD}), to identify modes with smaller energies, which have a chance to play
the role of the GS in the manifold of \textit{p}-wave states. The comparison
demonstrates that the CDD and CVD modes, represented by FPs (\ref{DD}) and (%
\ref{VD}), may have a chance to realize degenerate (in terms of the energy)
GSs, while the ODD mode, which corresponds to FP (\ref{ODD}), definitely
represents an excited state.

The FPs found above are stationary solutions with three or two nonzero
amplitudes, out of the six which constitute the GA, and a single frequency.
In addition to them, a more general stationary solution of Eqs. (\ref{a}), (%
\ref{b}) can be found, with four nonvanishing amplitudes and two different
frequencies:%
\begin{equation}
a_{0}=b_{0}=0,\quad a_{\pm 1}(t)=a_{\pm }e^{-4ib^{2}t},\quad b_{\pm
1}(t)=\pm be^{-2i\left( a_{+}^{2}+a_{-}^{2}\right) t},  \label{ab}
\end{equation}%
where $a_{\pm }$ and $b$ are three arbitrary real constants. There is also a
mirror image of FP (\ref{ab}) produced by swapping $a\rightleftarrows b$.
The norms (\ref{N-N})\ corresponding to FP (\ref{ab}) are
\begin{equation}
N_{a}=a_{+}^{2}+a_{-}^{2},~N_{b}=2b^{2}.  \label{NN}
\end{equation}%
In the particular case of $a_{+}\equiv a,a_{-}=0,b=a/\sqrt{2}$, the FP (\ref%
{ab}) goes over into the above one (\ref{VD}), which represents the coaxial
vortex-dipole complex. Another particular case corresponds to
\begin{equation}
a_{+}=a_{-}=b.  \label{aab}
\end{equation}%
It represents to a complex built of two orthogonal dipoles oriented along
the $x$ and $y$ axes, as shown in Fig. \ref{fig1}(d).

The investigation of the stability of FP (\ref{ab}) against small
perturbations by means of the Bogoliubov - de Gennes equations is too
cumbersome to be performed analytically. On the other hand, the energy
stability can be compared to that of the above FPs, for a particular case of
FP (\ref{ab}) with norms (\ref{NN}) equal to those of the solutions obtained
above: $N_{a}=N_{b}=a^{2}$. For this purpose, FP (\ref{ab}) is taken as%
\begin{gather}
a_{+}^{2}+a_{-}^{2}=a^{2},  \label{+-} \\
a_{\pm 1}(t)=a_{\pm }e^{-2ia^{2}t}\quad b_{\pm 1}(t)=\pm \frac{a}{\sqrt{2}}%
e^{-2ia^{2}t},  \label{4M}
\end{gather}%
which includes FP (\ref{aab}), as a particular case. As follows from Eqs. (%
\ref{phiGA}) and (\ref{psiGA}), this FP is built of a dipole in the $\Psi $
field oriented along the $z$ axis, and a mixed vortex-antivortex state in
the $\Phi $ field. The calculation of the respective value of Hamiltonian (%
\ref{HHH}) for the four-amplitude (4A) FP (\ref{4M}) yields

\begin{equation}
H_{\mathrm{4A}}=2a^{4}.  \label{H4M}
\end{equation}%
Note that expression (\ref{H4M}) does not depend on ratio $a_{+}/a_{-}$, see
Eq. (\ref{+-}).
Thus, Eq. (\ref{H4M}) demonstrates \emph{very broad degeneracy} of the GS,
in the framework of the GA: the same minimum of $H$ is realized by the FPs
given by Eqs. (\ref{DD}), (\ref{VD}), and (\ref{4M}), including the \emph{%
continuous degeneracy} in the 4A manifold with respect to the arbitrary
value of parameter $a_{+}/a_{-}$.

Finally, simulations of the coupled GPEs (\ref{phi}) and (\ref{psi}) in
imaginary time, which is a well-known numerical method for constructing
stationary states of GPEs \cite{Tosi}, made it easy to produce stationary
solutions, which are stable in real time too, and almost exactly correspond
to the FPs of the coaxial dipole-dipole and vortex-dipole types, as
predicted by the FPs (\ref{DD}) and (\ref{VD}), respectively. Shapes of
these solutions are very close to those displayed, in terms of \textit{Ans%
\"{a}tze} (\ref{phiGA}), (\ref{psiGA}), in the left and right columns of
Fig. \ref{fig1}. In the regimes of moderate values of Norm where GA with the leading terms works well, the difference in measured values of Norms between the FPs predicted by GA and direct numerical stationary solutions is within few percents.  On the other hand, stationary solutions close to the
orthogonal-dipole mode, corresponding to FP (\ref{ODD}), could not be
obtained by means of the imaginary- (or real-) time simulations. This
negative result is not surprising, as such methods cannot produce excited
states whose energies exceeds the GS energy, as shown in the present case by
Eq. (\ref{HOD}).

It is relevant to mention that evident solutions with identical coaxial
\textit{p}-waves (vortices) in components $\Phi $ and $\Psi $, both taken as
per Eq. (\ref{vort}), cannot be produced by the GA adopted above (because
axes of the respective sets of the \textit{p}-waves were chosen to be
mutually orthogonal, rather than parallel). For this reason, the coaxial
vortices are not considered here, but it is obvious that they are tantamount
to vortex states in the single GPE with the HO trapping potential and
self-repulsive nonlinearity, which were studied in detail before \cite%
{vortices}.

Concerning the absolute GS of the model based on the coupled GPEs, with
Hamiltonian (\ref{H}), additional calculations (perturbative analytical and
numerical) demonstrate that the obvious isotropic \textit{s}-wave solution,
with $l=0$, namely, $\Phi =\Psi =\exp \left( -i\mu t\right) u(r)$ (in the
case of equal norms of the two components, $N_{\Phi }=N_{\Psi }$), remains
the state which realizes the minimum of the energy for given norms.%

\section{Numerical results: Persistent time-periodic nonlinear dynamics}

\label{numres}

Numerical solutions of the coupled GPEs which generate the
(energy-degenerate) GSs of the $l=1$ manifold are presented in the previous
section. Here, our objective is to explore generic dynamical regimes
produced by the coupled GPE system and, in parallel, by its GA counterpart.

\subsection{The interaction of obliquely oriented dipoles}

As said above, pairs of mutually parallel or orthogonal dipoles, which
correspond to FPs (\ref{DD}) or (\ref{ODD}), respectively, form stationary
states with the shapes displayed in the left and middle columns of Fig. \ref%
{fig1} (recall that, while both these FPs are stable against small
perturbations, the stationary solution of the coupled GPEs corresponding to
FP (\ref{ODD}) is difficult to find in the numerics as it corresponds to an
excited state, as per Eq. (\ref{HOD})).

\begin{figure}[tbh]
\centering
\includegraphics[width=9cm]{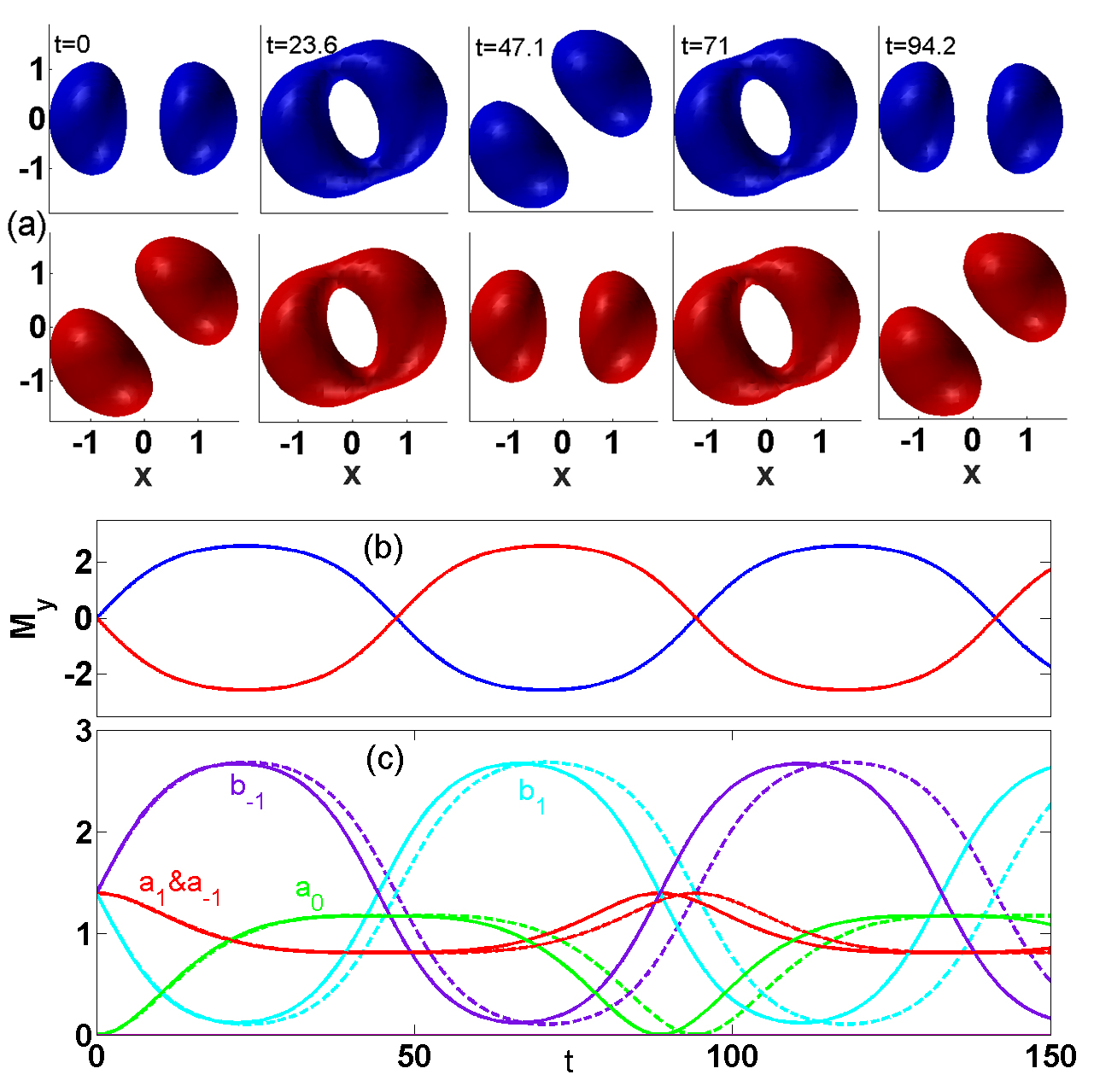} .
\caption{(Color online) Periodic oscillations of the pair of dipoles with
angle $\protect\theta =40^{\circ }$ between their orientations, initiated by
input (\protect\ref{inputDD}), with norms $N_{\Phi ,\Psi }=5.5$. (a)
Snapshots of the density isosurfaces of the two components, taken at moments
of time indicated in the panels, within one period of the oscillations ($%
T\approx 94$). (b) The evolution of the $y$-component of the angular momenta
of the two components, defined as per Eq. (\protect\ref{M}). (c) The
evolution of the amplitudes of the GA, based on \textit{Ans\"{a}tze} (%
\protect\ref{phiGA}), (\protect\ref{psiGA}), as produced by simulations of
Eqs. (\protect\ref{a}), (\protect\ref{b}) with initial conditions
corresponding to Eq. (\protect\ref{inputDD}) (solid curves), and the
evolution of their counterparts produced by the projection of the
numerically found solution of the coupled GPEs onto the Galerkin basis
(dashed curves). }
\label{fig2}
\end{figure}

The evolution of a pair of dipoles with mutual orientation different from 0
or 90 degrees is initiated by the input%
\begin{equation}
\Phi \left( \mathbf{r},t=0\right) =Ax\exp \left( -r^{2}/2\right) ,~\Psi
\left( \mathbf{r},t=0\right) =Ax^{\prime }\exp \left( -r^{2}/2\right) ,
\label{inputDD}
\end{equation}%
with amplitude $A$ and oblique angle $\theta $ between axes $x$ and $%
x^{\prime }$. Typical results are displayed in Fig. \ref{fig2} for $\theta
=40^{\circ }$ and a norm (\ref{N-N}) of each component of $N_{\Phi }=N_{\Psi
}=5.5$. Simulations of both the coupled GPEs (\ref{phi}), (\ref{psi}), and
of GA equations (\ref{a}), (\ref{b}) reveal periodic rocking oscillations of
the two dipoles, as illustrated in Fig. \ref{fig2}(b) by plots showing the
exchange of the $y$-component of the angular momentum (the one which is
different from zero in the \ present case) between the two wave-function
components ($\Phi $ and $\Psi $), see Eq. (\ref{M}). The oscillations
exhibit periodic transformations of the two components from the obliquely
oriented dipoles into coaxial vortices and back, as shown in Fig. \ref{fig2}%
(a), where the periodically emerging parallel vortices (in both components)
are oriented along the $y$-axis. The period of the oscillations is $T=94$ in
this case.

As mentioned above, pure vortical modes with $l=1$ have the absolute value
of their angular momentum equal to the norm, $M=N$. In the present case, the
largest value of the angular momentum, attained in the vortex-like
configurations at $t=23.6$ and $71$ in Fig. \ref{fig2}(a), is $\left(
\left\vert \left( M_{y}\right) _{\Phi ,\Psi }\right\vert \right) _{\max
}~=2.3$, while, as said above, the norms are $N_{\Phi ,\Psi }=5.5$, which
implies that the periodic conversion of the wave-field configurations into
the vortices is incomplete. Additional simulations demonstrate that the
ratio $\left( \left\vert \left( M_{y}\right) _{\Phi ,\Psi }\right\vert
\right) _{\max }/N_{\Phi ,\Psi }$ increases, following the growth of the
norms, i.e., the conversion into the vortices is more complete for a more
nonlinear system.

Finally, Fig.~\ref{fig2}(c) clearly shows that the GA provides a rather
accurate fit to the simulations of the full GPE system. In this connection,
additional computations demonstrate that the share of the total norm carried
by components of the full numerical solution with $l\geq 2$, which are
omitted in the framework of the GA based on Eqs. (\ref{phiGA}), (\ref{psiGA}%
), remains $\lesssim 2\%$ in the case of $N_{\Phi }=N_{\Psi }=5.5$, thus
justifying the application of the GA. Stronger nonlinearity, i.e., larger
values of the total norm, gradually leads to an increase of the discrepancy
of the full numerical solution from the GA (as an illustration, see Fig. \ref%
{fig3}(a) below, which shows the discrepancy in the oscillation period for a
much stronger nonlinearity).

\begin{figure}[tbh]
\centering
\includegraphics[width=9cm]{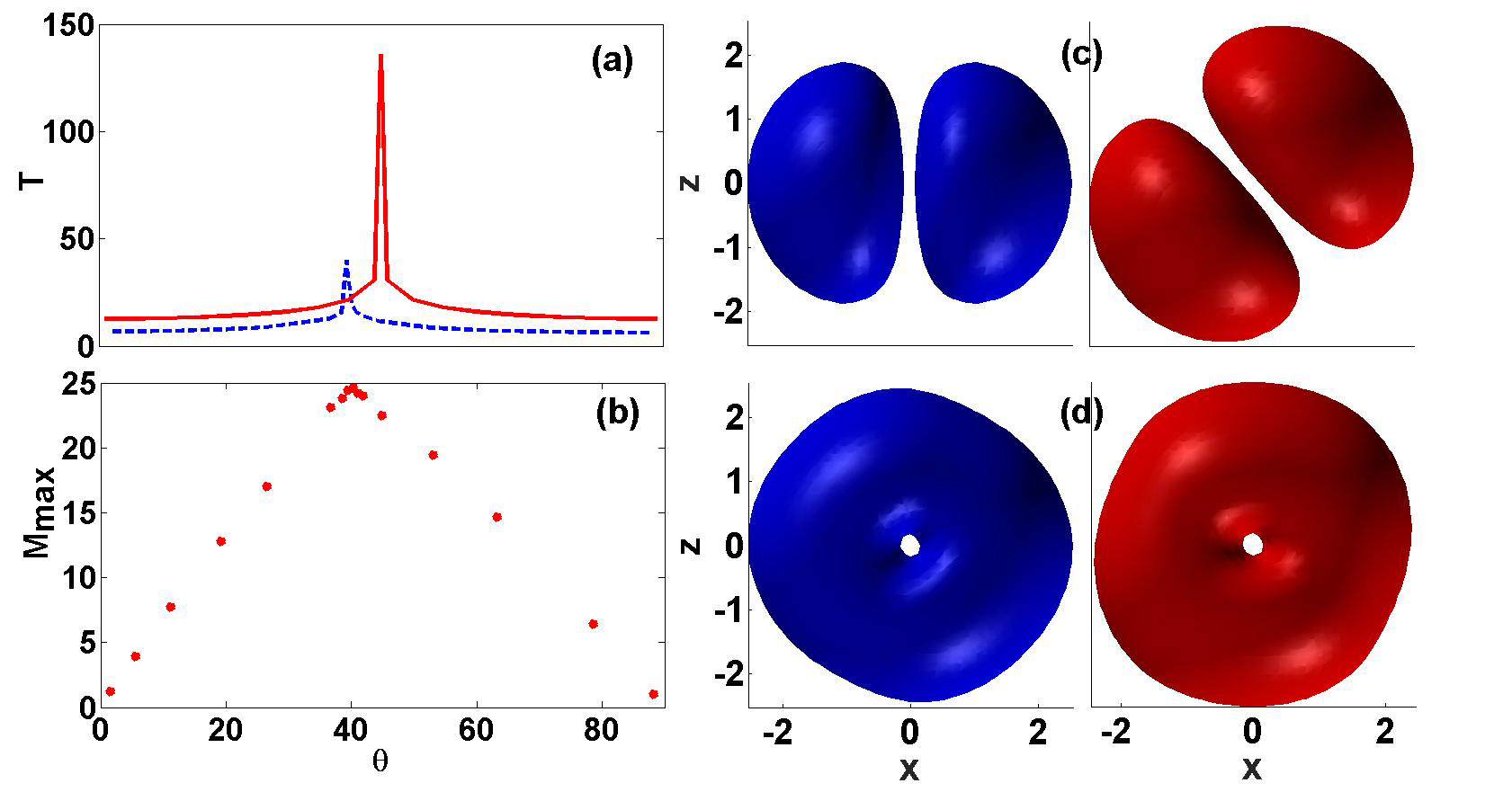} .
\caption{(Color online) Systematic results for the oscillatory dynamical
regime, initiated by input (\protect\ref{inputDD}), with initial angle $%
\protect\theta $ between orientations of the two dipoles, and norms $%
N_{\Phi}=N_{\Psi }=22.5$ (which exceed the norms in the case shown in Fig.
\protect\ref{fig2} \ by a factor of $5$, making the nonlinearity stronger
accordingly). (a) The dependence of the oscillation period on $\protect%
\theta $. The dashed blue and solid red plots display the results produced,
respectively, by direct simulations of the coupled GPEs (\protect\ref{phi})
and (\protect\ref{psi}), and by the Galerkin approximation, i.e., by
simulations of Eqs. (\protect\ref{a}), (\protect\ref{b}). (b) The largest
absolute value of the angular momentum, $M_{max}\equiv \left( \left\vert
\left( M_{y}\right) _{\Phi ,\Psi }\right\vert \right) _{\max }$, attained in
the course of the periodic evolution, versus $\protect\theta $. (c) and (d)
Snapshots of the density isosurfaces of the two wave-function components, $%
\left\vert \Phi \right\vert ^{2}$ and $\left\vert \Psi \right\vert ^{2}$, in
the configurations corresponding, respectively, to the input set (at $t=0$)
of two dipoles with $\protect\theta =40^{\circ }$, and to the most complete
conversion into the pair of coaxial vortices (at $t=5$), with the common
axis oriented along $y$.}
\label{fig3}
\end{figure}

The dependence of the basic features of the dynamical regime on initial
angle $\theta $ between the orientations of the two dipoles in the input
configuration (\ref{inputDD}) is displayed in Fig.~\ref{fig3} for a system
with stronger nonlinearity, measured by the norms, $N_{\Phi ,\Psi }=22.5$,
which are larger by a factor of $5$ than the case shown in Fig. \ref{fig2}.
The results of Figs.~\ref{fig3}(a) and (b) show a strong dependence of the
oscillation period and of the efficiency of the periodic conversion of the
wave-function configuration into the set of coaxial vortices, which is
measured by the above-mentioned ratio $\left( \left\vert \left( M_{y}\right)
_{\Phi ,\Psi }\right\vert \right) _{\max }/N_{\Phi ,\Psi }$, on the initial
mutual-orientation angle $\theta $. It is seen that both $T$ and $\left(
\left\vert \left( M_{y}\right) _{\Phi ,\Psi }\right\vert \right) _{\max
}/N_{\Phi ,\Psi }$ attain a sharp maximum at $\theta \approx 40^{\circ }$
(the GA predicts the maximum of $T$ exactly at $\theta =45^{\circ }$, with a
symmetric dependence of $T$ on $\left( \theta -45^{\circ }\right) $, in Fig. %
\ref{fig3}(a)). In fact, the difference between the full GPE dynamics and
its GA-truncated version is largest precisely at $\theta =45^{\circ }$,
which explains the large discrepancy in the maximum values of the GPE- and
GA-predicted periods, observed in Fig. \ref{fig3}(a). Note also that the
largest value of the momentum may even exceed the norm, \textit{viz}.,$%
\left( \left\vert \left( M_{y}\right) _{\Phi ,\Psi }\right\vert \right)
_{\max }/N_{\Phi ,\Psi }\approx 1.11$ at $\theta =40^{\circ }$, as can be
seen in Fig. \ref{fig3}(b).

In this connection, it is relevant to mention that the dynamical equations
of the GA (\ref{a}) and (\ref{b}) feature an exact scaling invariance
(unlike the GPEs (\ref{phi}), (\ref{psi}), in which the scaling invariance
is broken by fixing the HO length to be $1$; in that case, the precession
and nutation periods are more complex functions of $1/N$, as found for
cross-vortices in Ref. \cite{arxiv}). In the case of $N_{\Phi
}=N_{\Psi}\equiv N/2$, this implies that the oscillation period scales as $%
T\sim 1/N$.
Accordingly, the increase of the norm by the factor of $5$ in the case
displayed in Fig. \ref{fig3} for $\theta =40^{\circ }$, in comparison with
the case of Fig. \ref{fig2}, leads to the reduction of the GA-predicted
period from $T=94$ in Fig. \ref{fig2} to $T\approx 19$ in Fig.~\ref{fig3}(a).

\subsection{The interaction of mutually orthogonal dipole and vortex}

The analysis of the FPs presented in the previous section has produced
stable stationary solutions for the coaxial vortex-dipole complex, which
corresponds to FP (\ref{VD}) and the right column in Fig. \ref{fig1}. Below
we consider the dynamics initiated by an input composed of mutually
perpendicular vortex in one component and dipole in the other, which are
oriented in the $z$ and $y$ directions, respectively%
\begin{equation}
\Phi \left( \mathbf{r},t=0\right) =A\left( x+iy\right) \exp \left(
-r^{2}/2\right) ,~\Psi \left( \mathbf{r},t=0\right) =Ax\exp \left(
-r^{2}/2\right) ,  \label{inputVD}
\end{equation}%
cf. Eq. (\ref{inputDD}).

\begin{figure}[tbh]
\centering
\includegraphics[width=9cm]{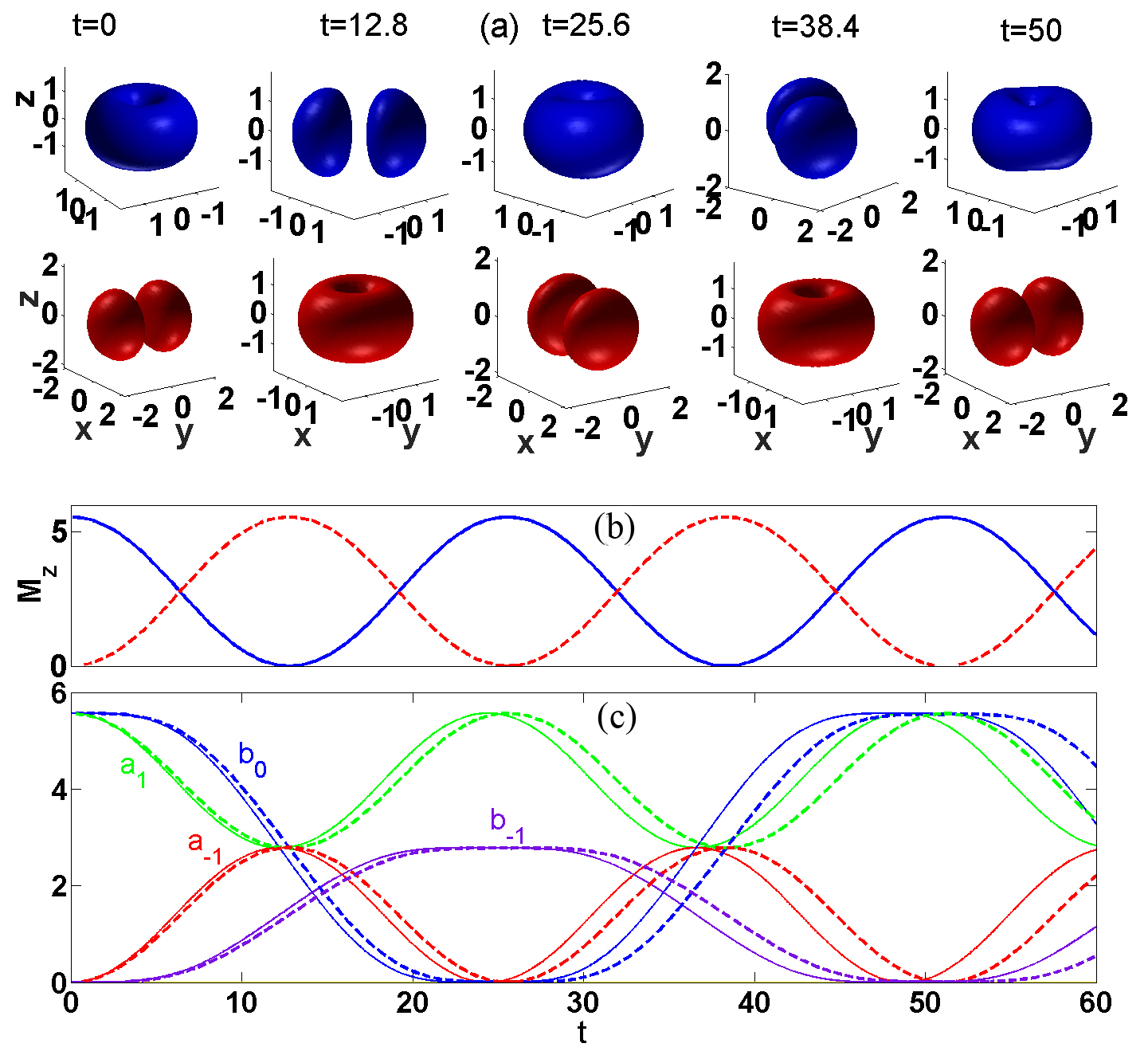} .
\caption{(Color online) (a) A set of snapshots of the density profiles of
the two components illustrating the periodic evolution initiated by input (%
\protect\ref{inputVD}) composed of mutually perpendicular vortex and dipole,
with norms $N_{\Phi }=N_{\Psi }=5.5$. (b) The time dependence of the angular
momenta, $\left( M_{z}\right) _{\Phi ,\Psi }$, of the two components (blue
solid and red dashed curves correspond to the components which represent the
vortex and the dipole at $t=0$, respectively). (c) The evolution of the
amplitudes of the eigenfunctions on which the Galerkin approximation (GA) is
based. The result of simulations of the GA equations (\protect\ref{a}), (%
\protect\ref{b}) is shown by solid curves. Dashed curves display the results
produced by the projection of the numerical solution of the full GPE system (%
\protect\ref{phi}), (\protect\ref{psi}) onto the GA modal basis.}
\label{fig4}
\end{figure}

Simulations of these configurations exhibit exchange of the angular momentum
between the two wave functions. Initially, the vortex donates its $z $%
-component of the angular momentum to the dipole, which initially carries no
angular momentum. As a result, the dipole starts rotating around the $z$
axis. After a quarter of the period of the resulting oscillations (at $%
t=12.8 $), the dipole absorbs all the angular momentum from the vortex,
itself becoming a fully shaped vortex, which features the above-mentioned
respective condition, $\left\vert M_{z}\right\vert =N$. Simultaneously, the
original vortex component, devoid of the angular momentum, acquires a
dipole-mode shape. These metamorphoses are clearly seen in the second column
of snapshots displayed Fig. \ref{fig4}(a). At the time corresponding to a
half-period, $t=25.6$, the wave functions return to their original shapes,
with the dipole rotated by 90 degrees relative to its original direction
(the third column in Fig. \ref{fig4}(a)). A nearly perfect recovery of the
initial configuration in both components is observed at the end of the full
period, $t=50$. Fig.~\ref{fig2}(b) tracks the periodic exchange of the
angular momentum between the two wave functions, while Fig.~\ref{fig2}(c)
demonstrates the evolution of the amplitudes of the eigenfunctions, the
projection onto which corresponds to the GA. Close agreement between the GA
results and their counterparts generated by simulations of the full GPE
system is obtained again.

\subsection{The interaction of non-coaxial vortices}

\begin{figure}[tbh]
\centering
\includegraphics[width=9cm]{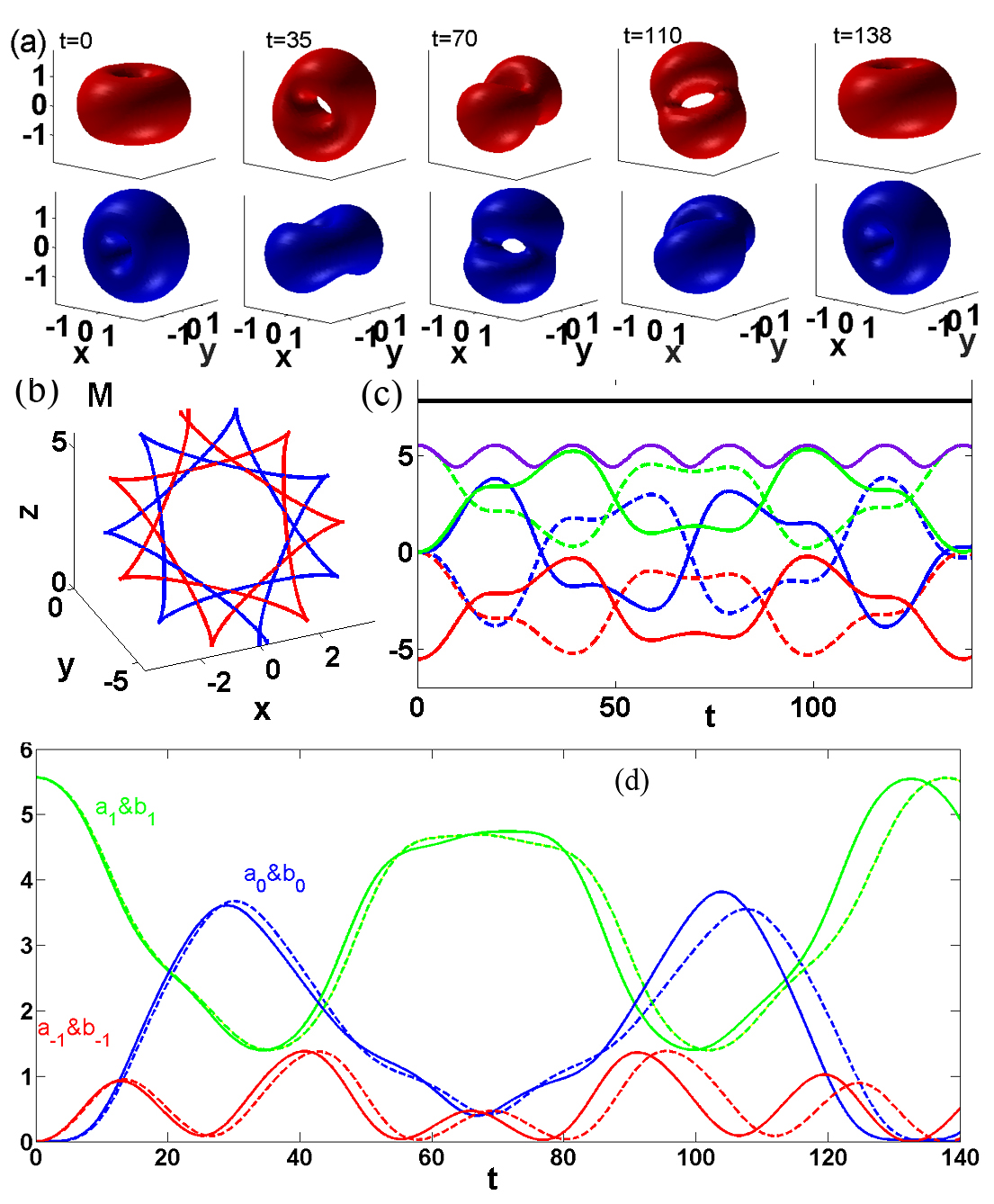} .
\caption{(Color online) The evolution of the binary system initiated by
input (\protect\ref{inputVV}) composed of two vortices, with the initial
angular-momenta vectors directed along $z$ and $y$, respectively, and norms $%
N_{\Phi ,\Psi }=5.56$. (a) Snapshots of density isosurfaces of the two
components at time moments indicated in panels, within an oscillation
period, $T=138$. (b) Trajectories of end points of the 3D angular-momentum
vectors of the two components (the red and blue colors\ correspond to ones
in (a)). (c) Green, red, and blue lines render, severally, the corresponding
evolution of the $z$, $y$, and $x$ components of the angular momenta of the
two wave function (the dashed and solid lines correspond, respectively, to
the $\Phi $ and $\Psi $ wave functions). The violet curve represents the
evolution of the total angular momentum of each condensate, while the
horizontal black line shows the sum of the two angular momenta which remains
constant. (d) The corresponding evolution of the GA amplitudes (defined by
\textit{Ans\"{a}tze }(\protect\ref{phiGA}) and (\protect\ref{psiGA})), as
produced by simulations of Galerkin equations (\protect\ref{a}), (\protect
\ref{b}), and by simulations of the full GPEs (\protect\ref{phi}), (\protect
\ref{psi}), is depicted by solid and dashed lines, respectively.}
\label{fig5}
\end{figure}

The analysis of the FPs performed in the previous section did not produce
any stationary state composed of two non-coaxial vortices. Here, we explore
the dynamics initiated by a pair of identical vortices with mutually
perpendicular orientations, aligned with the $z$- and $y$-axes:%
\begin{equation}
\Phi \left( \mathbf{r},t=0\right) =A\left( x+iy\right) \exp \left(
-r^{2}/2\right) ,~\Psi \left( \mathbf{r},t=0\right) =A\left( z+ix\right)
\exp \left( -r^{2}/2\right) ,  \label{inputVV}
\end{equation}%
fixing the amplitude as $A=1$ and norms as $N_{\Phi}=N_\Psi=5.56$ (this case
of non-coaxial vortices was studied in~\cite{arxiv}, however, for a case of
dominating self-phase interactions). The resulting evolution of the binary
system is displayed in Fig.~\ref{fig5}, which shows that, unlike the
interacting dipoles (cf. Figs. \ref{fig2} and \ref{fig4}), the motion does
not amount to rotation about a particular axis. Instead, the vortices
undergo complex deformation, related to periodic generation of all the three
components of the angular momentum in each wave function, as shown in Fig. %
\ref{fig5}(c) (the net angular momentum, defined as per Eq. (\ref{M}),
remains conserved). The evolution of the angular momenta of each vortex is
shown in vectorial (Fig. \ref{fig5}(b)) and scalar (Fig. \ref{fig5}(c))
representations. The vectorial representation (Fig. \ref{fig5}(b))
demonstrates the motion of the locus of each vortex axis in 3D space. The
red trajectory pertains to the vortex initially aligned with the $z$-axis,
while the blue trajectory corresponds to the initially $y$-oriented vortex.
The trajectories features precession with nutations qualitatively similar to
the case of mixed inter- and intra-species interactions reported in \cite%
{arxiv}. Scalar representation (Fig.~\ref{fig5}(c)) separates each component
of the angular momentum and shows its evolution. Green curves solid and
dashed pertain to the $z$-component of the angular momenta of each vortex
respectively. Blue curves (solid and dashed) represent the evolution of the $%
x$-components of the angular momenta of the two vortices and finally the red
curves (solid and dashed) represent the evolution of the $y$-components of
the angular momenta for each of the vortices. Moreover, we present the
evolution of the total angular momenta of both condensates which perfectly
overlap by the violet curve. This violet curve demonstrates that in the case
of the two vortices unlike the case of vortex-dipole we do not have an
exchange of angular momenta between the two fields. The horizontal black
line in Fig.~\ref{fig5}(c) shows the sum of the two angular momenta which is
an integral of motion. As above, Fig.~\ref{fig5}(d) demonstrates that the GA
provides an accurate description of the present dynamical regime in terms of
the finite-mode truncation.

Lastly, the analysis was also performed for inputs similar to the one
defined by Eq.~(\ref{inputVV}), but with the angle between the
angular-momentum vectors of the two vortices different from $90^{\circ }$.
These results (not shown) are similar to those presented here. In
particular, the evolution of the two angular momenta in the vectorial form
again exhibits precession-like motion combined with nutations.

\section{Conclusion}

\label{sum}

We have introduced a system of two 3D fields, trapped in the HO potential,
which interact through the repulsive cubic nonlinearity. The system can be
implemented in the form of a binary BEC, with the interaction dominated by
the inter-component repulsion, that can be enhanced by means of the Feshbach
resonance. Parallel to the system of two coupled GPEs (Gross-Pitaevskii
equations), we have introduced a finite-mode truncation, which amounts to
the six-mode GA (Galerkin approximation), based on the triplets of HO
eigenstates in each component, with quantum numbers $l=1$ and $m=1,0,-1$.
The first result is that FPs (fixed points) of the GA almost exactly predict
the GS (ground-state) manifold for $l=1$, which features unusually broad
degeneracy, including complexes in the form of a dipole coaxial with a
vortex or another dipole. Dynamical regimes were initiated by inputs built
as pairs of non-coaxial dipoles and/or vortices, including pairs of
orthogonally oriented vortices. As a result, the system gives rise to stable
dynamics, characterized by periodic conversions between dipoles and
vortices. In this case too, results produced by the GA are well corroborated
by direct simulations of the coupled GPEs.

As a development of the present system, it is relevant to expand it to a
three-component spinor model, in which inter-component interactions may
include four-wave mixing, in addition to the mutual repulsion \cite{spinor}.
In particular, a natural possibility will be to consider the interaction
between three mutually orthogonal vortices in the three-component system
trapped in a common isotropic 3D HO potential. On the other hand, the high
accuracy produced by the GA in the present system suggests that this
approximation may be used as an efficient tool for the study of complex
stationary states and dynamical regimes in other 3D classical-field models.

\begin{acknowledgements}
RD and TM acknowledge support from the Deutsche Forschungsgemeinschaft (DFG) via GRK 1464, and computation time provided by the Paderborn Center for Parallel Computing (PC$^2$). VVK acknowledges support from FCT (Portugal) under grant UID/FIS/00618/2013. BAM appreciated hospitality of the Department of Physics at the University of Paderborn.
\end{acknowledgements}


\begin{thebibliography}{99}
\bibitem{vortex-rings} M. S. Volkov and E. Wohnert, Spinning Q-balls, Phys.
Rev. D \textbf{66}, 085003 (2002); B. Kleihaus, J. Kunz, and Y. Shnir,
Monopoles, antimonopoles, and vortex rings, \textit{ibid}. \textbf{68},
101701 (2003); B. Kleihaus, J. Kunz, and Y. Shnir, Monopole-antimonopole
chains and vortex rings, \textit{ibid}. \textbf{70}, 065010 (2004);
Gravitating monopole-antimonopole chains and vortex rings, \textit{ibid}.
\textbf{71}, 024013 (2005); J. Kunz, U. Neemann, and Y. Shnir,Transitions
between vortex rings, and monopole--antimonopole chains, Phys. Lett. B
\textbf{640}, 57 (2006); E. Radu and M. S. Volkov, Spinning electroweak
sphalerons, Phys. Rev. D \textbf{79}, 065021 (2009); J. Garaud, E. Radu, and
M. S. Volkov, Stable cosmic vortons, Phys. Rev. Lett. \textbf{111}, 171602
(2013).

\bibitem{knots} L. Faddeev and A. J. Niemi, Stable knot-like structures in
classical field theory, Nature (London) \textbf{387}, 58 (1997); Partially
dual variables in SU(2) Yang-Mills theory, Phys. Rev. Lett. \textbf{82},
1624 (1999); E. Babaev, L. D. Faddeev, and A. J. Niemi, Hidden symmetry and
knot solitons in a charged two-condensate Bose system, Phys. Rev. B \textbf{%
65}, 100512(R) (2002); J. J\"{a}ykk\"{a} and J. M. Speight, Supercurrent
coupling destabilizes knot solitons, Phys. Rev. D \textbf{84}, 125035 (2011).

\bibitem{hopfions} J. Gladikowski and M. Hellmund, Static solitons with
nonzero Hopf number, Phys. Rev. D \textbf{56}, 5194 (1997); H. Aratyn, L. A.
Ferreira, and A. H. Zimerman, Exact static soliton solutions of
(3+1)-dimensional integrable theory with nonzero Hopf numbers, Phys. Rev.
Lett. \textbf{83}, 723 (1999); J. J\"{a}ykk\"{a}, J. Hietarinta, and P.
Salo, Topologically nontrivial configurations associated with Hopf charges
investigated in the two-component Ginzburg-Landau model, Phys. Rev. B
\textbf{77}, 094509 (2008); J. M. Speight, Supercurrent coupling in the
Faddeev--Skyrme model, J. Geom. Phys. \textbf{60}, 599 (2010).

\bibitem{skyrmions} M. F. Atiyah and N. S. Manton, Phys. Lett. B \textbf{222}%
, 438-442 (1989); C. J. Houghton, N. S. Manton, and P. M. Sutcliffe,
Rational maps, monopoles and skyrmions, Nucl. Phys. B \textbf{510}, 507-537
(1998).


\bibitem{review} E. Radu and M. S. Volkov, Stationary ring solitons in field
theory - Knots and vortons, Phys. Rep. \textbf{468}, 101 (2008).

\bibitem{hadrons} M. Bender, P. H. Heenen, and P. G. Reinhard,
Self-consistent mean-field models for nuclear structure, Rev. Mod. Phys.
\textbf{75}, 121 (2003); T. Sakai, and S. Sugimoto, Low energy hadron
physics in holographic QCD, Prog. Theor. Phys. \textbf{113}, 843-882 (2005);
A. Pomarol and A. Wulzer, Baryon physics in holographic QCD, Nucl. Phys.
\textbf{809}, 347-361 (2009).

\bibitem{ferro} N. R. Cooper, Propagating magnetic vortex rings in
ferromagnets, Phys. Rev. Lett. \textbf{82}, 1554 (1999); S. Seki, X. Z. Yu,
S. Ishiwata, and Y. Tokura, Observation of skyrmions in a multiferroic
material, Science \textbf{336}, 198-201 (2012); A. Fert,V. Cros, and J.
Sampaio, Skyrmions on the track, Nature Nanotechnology \textbf{8}, 152-156
(2013); Y. Nii, T. Nakajima, A. Kikkawa, Y. Yamasaki, K. Ohishi, J. Suzuki,
Y. Taguchi, T. Arima, Y. Tokura, and  Y. Iwasa, Uniaxial stress control of
skyrmion phase, Nature Commun. \textbf{6}, 8539 (2015); Y. Nahas, S.
Prokhorenko, L. Louis, Z. Gui, I. Kornev, and L. Bellaiche, Discovery of
stable skyrmionic state in ferroelectric nanocomposites, \textit{ibid}.
\textbf{6}, 8542 (2015).

\bibitem{super} E. Babaev, Dual neutral variables and knot solitons in
triplet superconductors, Phys. Rev. Lett. \textbf{88}, 177002 (2002);
Non-Meissner electrodynamics and knotted solitons in two-component
superconductors, Phys. Rev. B \textbf{79}, 104506 (2009).

\bibitem{semi} A. Neubauer, C. Pfleiderer, B. Binz, A. Rosch, R. Ritz, P. G.
Niklowitz, and P. B\"{o}ni, Topological Hall effect in the A phase of MnSi,
Phys. Rev. Lett. \textbf{102}, 186602 (2009); I. Kezsmarki, S. Bordacs, P.
Milde, E. Neuber, L. M. Eng, J. S. White, H. M. Ronnow, C. D. Dewhurst, M.
Mochizuki, K. Yanai, H. Nakamura, D. Ehlers, V. Tsurkan, and A. Loidl,
Neel-type skyrmion lattice with confined orientation in the polar magnetic
semiconductor GaV4S8, Nature Materials \textbf{14,} 1116 (2015).

\bibitem{BEC-skyrmions} J. Ruostekoski and J. R. Anglin, Creating vortex
rings and three-dimensional skyrmions in Bose-Einstein condensates, Phys.
Rev. Lett. \textbf{86}, 3934 (2001); R. A. Battye, N. R. Cooper and P. M.
Sutcliffe, Stable skyrmions in two-component Bose-Einstein condensates,
\textit{ibid}. \textbf{88}, 080401 (2002); C. M. Savage and J. Ruostekoski,
Energetically stable particlelike skyrmions in a trapped Bose-Einstein
condensate, \textit{ibid}. \textbf{91}, 010403 (2003); J. Ruostekoski, and
J. R. Anglin, Monopole core instability and Alice rings in spinor
Bose-Einstein condensates, \textit{ibid}. \textbf{91}, 190402 (2003); C.-H.
Hsueh, S.-C. Gou, T.-L. Horng, and Y.-M. Kao, Vortex-ring solutions of the
Gross-Pitaevskii equation for an axisymmtrically trapped Bose-Einstein
condensate, J. Phys. B: At. Mol. Opt. Phys. \textbf{40}, 4561-4571 (2007);
M. Kan\'{a}sz-Nagy, B. D\'{o}ra, E. A. Demler, and G. Zar\'{a}nd,
Stabilizing the false vacuum: Mott skyrmions. Sci. Rep. \textbf{5}, 7692
(2014); T. Kaneda and H. Saito, Collision dynamics of skyrmions in a
two-component Bose-Einstein condensate, Phys. Rev. A \textbf{93}, 033611
(2016).

\bibitem{skyrmion-exper} L. S. Leslie, A. Hansen, K. C. Wright, B. M.
Deutsch, and N. P. Bigelow, Creation and detection of skyrmions in a
Bose-Einstein condensate, Phys. Rev. Lett. \textbf{103}, 250401 (2009); J.
Y. Choi, W. J. Kwon, and Y. I. Shin, Observation of topologically stable 2D
skyrmions in an antiferromagnetic spinor Bose-Einstein condensate, \textit{%
ibid}. \textbf{108}, 035301 (2012).

\bibitem{BEC-knots} Y.-K. Liu, S. Feng, and S.-J. Yang, Stable knots in the
trapped Bose-Einstein condensates, EPL \textbf{106}, 50005 (2014).

\bibitem{BEC-hopfion} Y. M. Bidasyuk, A. V. Chumachenko, O. O. Prikhodko, S.
I. Vilchinskii, M. Weyrauch, and A. I. Yakimenko, Stable Hopf solitons in
rotating Bose-Einstein condensates, Phys. Rev. A \textbf{92}, 053603 (2015);
R. N. Bisset, W. Wang, C. Ticknor, R. Carretero-Gonz\'{a}lez, D. J.
Frantzeskakis, L. A. Collins, and P. G. Kevrekidis, Robust vortex lines,
vortex rings, and hopfions in three-dimensional Bose-Einstein condensates,
Phys. Rev. A \textbf{92}, 063611 (2015).

\bibitem{monopole-exper} M. W. Ray, E. Ruokokoski, S. Kandel, M. M\"{o}tt%
\"{o}nen, and D. S. Hall, Observation of Dirac monopoles in a synthetic
magnetic field, Nature \textbf{505}, 657-660 (2014); M. W. Ray, E.
Ruokokoski, K. Tiurev, M. M\"{o}tt\"{o}nen, and D. S. Hall, Observation of
isolated monopoles in a quantum field, Science \textbf{348}, 544-547 (2015).

\bibitem{general-reviews} Yu. S. Kivshar and D. E. Pelinovsky, Self-Focusing
and Transverse Instabilities of Solitary Waves, Phys. Rep. \textbf{331}, 117
(2000); Y. S. Kivshar and G. P. Agrawal, \textit{Optical Solitons: From
Fibers to Photonic Crystals} (Academic Press, San Diego, 2003); B. A.
Malomed, D. Mihalache, F. Wise, and L. Torner, Spatiotemporal optical
solitons, J. Optics B: Quant. Semicl. Opt. \textbf{7}, R53-R72 (2005); A. S.
Desyatnikov, L. Torner, and Y. S. Kivshar, Optical Vortices and Vortex
Solitons, Progr. Opt. \textbf{47}, 1 (2005); D. Mihalache, Linear and
nonlinear light bullets: recent theoretical and experimental studies, Rom.
J. Phys. \textbf{57}, 352-371 (2012); V. S. Bagnato, D. J. Frantzeskakis, P.
G. Kevrekidis, B. A. Malomed, and D. Mihalache, Bose-Einstein condensation:
Twenty years after, Rom. Rep. Phys. \textbf{67}, 5-50 (2015).

\bibitem{RMP} Y. V. Kartashov, B. A. Malomed, and L. Torner, Solitons in
nonlinear lattices, Rev. Mod. Phys. \textbf{83}, 247-306 (2011).

\bibitem{Wagner} R. McLeod, K. Wagner, and S. Blair, (3+1)-dimensional
optical soliton dragging logic, Phys. Rev. A \textbf{52}, 3254 (1995).

\bibitem{interferometry} A. D. Martin and J. Ruostekoski, Quantum dynamics
of atomic bright solitons under splitting and recollision, and implications
for interferometry, New J. Phys. \textbf{14}, 043040 (2012); J. Cuevas, P.
G. Kevrekidis, B. A. Malomed, P. Dyke, and R. G. Hulet, Interactions of
solitons with a Gaussian barrier: Splitting and recombination in quasi-1D
and 3D, New J. Phys. \textbf{15}, 063006 (2013); J. H. V. Nguyen, P. Dyke,
D. Luo, B. A. Malomed, and R. G. Hulet, Collisions of matter-wave solitons,
Nature Phys. \textbf{10}, 918-922 (2014); G.\thinspace D. McDonald,
C.\thinspace C.\ N. Kuhn, K.\thinspace S. Hardman, S. Bennetts, P.\thinspace
J. Everitt, P.\thinspace A. Altin, J.\thinspace E. Debs, J.\thinspace D.
Close, and N.\thinspace P. Robins, Bright solitonic matter-wave
interferometer, Phys. Rev. Lett. \textbf{113}, 013002 (2014).

\bibitem{collapse} L. Berg\'{e}, Wave collapse in physics: principles and
applications to light and plasma waves, Phys. Rep. \textbf{303}, 259-370
(1998); E. A. Kuznetsov and F. Dias, Bifurcations of solitons and their
stability, Phys. Rep. \textbf{507}, 43-105 (2011).

\bibitem{lattice} N. K. Efremidis, S. Sears, D. N. Christodoulides, J. W.
Fleischer, and M. Segev, Discrete solitons in photorefractive optically
induced photonic lattices, Phys. Rev. E \textbf{66}, 046602 (2002); B. B.
Baizakov, B. A. Malomed, and M. Salerno, Multidimensional solitons in
periodic potentials, Europhys. Lett. \textbf{63}, 642 (2003); J. Yang and Z.
H. Musslimani, Fundamental and vortex solitons in a two-dimensional optical
lattice, Opt. Lett. \textbf{28}, 2094 (2003); D. Mihalache, D. Mazilu, F.
Lederer, Y. V. Kartashov, L.-C. Crasovan, and L. Torner, Stable
three-dimensional spatiotemporal solitons in a two-dimensional photonic
lattice, Phys. Rev. E \textbf{70}, 055603(R) (2004).

\bibitem{photorefr} D. Neshev, T. J. Alexander, E. A. Ostrovskaya, Y. S.
Kivshar, H. Martin, I. Makasyuk, and Z. Chen, Observation of discrete vortex
solitons in optically induced photonic lattices, Phys. Rev. Lett. \textbf{92}%
, 123903 (2004); J. W. Fleischer, G. Bartal, O. Cohen, O. Manela, M. Segev,
J. Hudock, and D. N. Christodoulides, Observation of vortex-ring
\textquotedblleft discrete\textquotedblright solitons in 2D photonic
lattices, \textit{ibid}. \textbf{92}, 123904 (2004).

\bibitem{Fukuoka} H. Sakaguchi, B. Li, and B. A. Malomed, Creation of
two-dimensional composite solitons in spin-orbit-coupled self-attractive
Bose-Einstein condensates in free space, Phys. Rev. E \textbf{89}, 032920
(2014).

\bibitem{HP} Y.-C. Zhang, Z.-W. Zhou, B. A. Malomed, and H. Pu, Stable
solitons in three dimensional free space without the ground state:
Self-trapped Bose-Einstein condensates with spin-orbit coupling, Phys. Rev.
Lett. \textbf{115}, 253902 (2015).

\bibitem{Pit} L. P. Pitaevskii and S. Stringari, \textit{Bose-Einstein
Condensation}, Oxford University Press (Oxford, 2003).

\bibitem{ICFO} O. V. Borovkova, Y. V. Kartashov, B. A. Malomed, and L.
Torner, Algebraic bright and vortex solitons in defocusing media, Opt. Lett.
\textbf{36}, 3088-3090 (2011); O. V. Borovkova, Y. V. Kartashov, L. Torner,
and B. A. Malomed, Bright solitons from defocusing nonlinearities, Phys.
Rev. E \textbf{84}, 035602 (R) (2011).

\bibitem{further} R. Driben, Y. V. Kartashov, B. A. Malomed, T. Meier, and
L. Torner, Soliton gyroscopes in media with spatially growing repulsive
nonlinearity, Phys. Rev. Lett. \textbf{112}, 020404 (2014); Y. V. Kartashov,
B. A. Malomed, Y. Shnir, and L. Torner, Twisted toroidal vortex-solitons in
inhomogeneous media with repulsive nonlinearity, \textit{ibid}. \textbf{113}%
, 264101 (2014); R. Driben, Y. Kartashov, B. A. Malomed, T. Meier, and L.
Torner, Three-dimensional hybrid vortex solitons, New J. Phys. \textbf{16},
063035 (2014); R. Driben, N. Dror, B. Malomed, and T. Meier, Multipoles and
vortex multiplets in multidimensional media with inhomogeneous defocusing
nonlinearity, \textit{ibid}. \textbf{17}, 083043 (2015); R. Driben, T.
Meier, and B. A. Malomed, Creation of vortices by torque in multidimensional
media with inhomogeneous defocusing nonlinearity, Sci. Rep. \textbf{5}, 9420
(2015).

\bibitem{vortices} S. Sinha, Semiclassical analysis of collective
excitations in Bose-Einstein condensate, Phys. Rev. A \textbf{55}, 4325
(1997); R. J. Dodd, K. Burnett, M. Edwards, and C. W. Clark, Excitation
spectroscopy of vortex states in dilute Bose-Einstein condensed gases,
\textit{ibid}. \textbf{56}, 587 (1997); T. Isoshima, M. Nakahara, T. Ohmi,
and K. Machida, Creation of a persistent current and vortex in a
Bose-Einstein condensate of alkali-metal atoms, \textit{ibid}. \textbf{61},
063610 (2000); A. A. Svidzinsky and A. L. Fetter, Stability of a vortex in a
trapped Bose-Einstein condensate, Phys. Rev. Lett. \textbf{84}, 5919 (2000);
V. M. Lashkin, Stable three-dimensional spatially modulated vortex solitons
in Bose-Einstein condensates, Phys. Rev. A \textbf{78}, 033603 (2008); T. P.
Simula, T. Mizushima, and K. Machida, Vortex waves in trapped Bose-Einstein
condensates, \textit{ibid}. \textbf{78}, 053604 (2008).

\bibitem{vort-clusters} L. C. Crasovan, G. Molina-Terriza, J. P. Torres, L.
Torner, V. M. P\'{e}rez-Garc\'{\i}a, and D. Mihalache, Globally linked
vortex clusters in trapped wave fields, Phys. Rev. E \textbf{66}, 036612
(2002); L. C. Crasovan, V. Vekslerchik, V. M. P\'{e}rez-Garc\'{\i}a, J. P.
Torres, D. Mihalache, and L. Torner, Stable vortex dipoles in nonrotating
Bose-Einstein condensates. Phys. Rev. A \textbf{68}, 063609 (2003); V. M.
Lashkin, Two-dimensional multisolitons and azimuthons in Bose-Einstein
condensates, \textit{ibid}. \textbf{77}, 025602 (2008); S. Middelkamp, P. J.
Torres, P. G. Kevrekidis, D. J. Frantzeskakis, R. Carretero-Gonz\'{a}lez, P.
Schmelcher, D. V. Freilich, and D. S. Hall, Guiding-center dynamics of
vortex dipoles in Bose-Einstein condensates, \textit{ibid}. \textbf{84},
011605 (R) (2011).

\bibitem{vort-exper} G. A. Swartzlander, Jr. and C. T. Law, Optical vortex
solitons observed in Kerr nonlinear media, Phys. Rev. Lett. \textbf{69,}
2503 (1992); M. R. Matthews, B. P. Anderson, P. C. Haljan, D. S. Hall, C. E.
Wieman, and E. A. Cornell, Vortices in a Bose-Einstein condensate, Phys.
Rev. Lett. \textbf{83,} 2498 (1999); B. P. Anderson, P. C. Haljan, C. A.
Regal, D. L. Feder, L. A. Collins, C. W. Clark, and E. A. Cornell, Watching
dark solitons decay into vortex rings in a Bose-Einstein condensate, \textit{%
ibid}. \textbf{86}, 2926 (2001); A. E. Leanhardt, A. G\"{o}rlitz, A. P.
Chikkatur, D. Kielpinski, Y. I. Shin, D. E. Pritchard, and W. Ketterle,
Imprinting vortices in a Bose-Einstein condensate using topological phases,
\textit{ibid}. \textbf{89}, 190403 (2002); V. Bretin, P. Rosenbusch, F.
Chevy, G. V. Shlyapnikov, and J. Dalibard, Quadrupole oscillation of a
single-vortex Bose-Einstein condensate: Evidence for Kelvin modes, \textit{%
ibid}. \textbf{90}, 100403 (2003).

\bibitem{vort-rev} A. L. Fetter, Rotating trapped Bose-Einstein condensates,
Rev. Mod. Phys. \textbf{81}, 647 (2009).

\bibitem{2comp-vort} J. J. Garc\'{\i}a-Ripoll and V. M. P\'{e}rez-Garc\'{\i}%
a, Stable and unstable vortices in multicomponent Bose-Einstein condensates,
Phys. Rev. Lett. \textbf{84}, 4264 (2000); K. Kasamatsu, M. Tsubota, and M.
Ueda, Vortices in multicomponent Bose-Einstein condensates, Int. J. Mod.
Phys. B \textbf{19}, 1835-1904 (2005); K. M. Mertes, J. W. Merrill, R.
Carretero-Gonz\'{a}lez, D. J. Frantzeskakis, P. G. Kevrekidis, and D. S.
Hall, Nonequilibrium dynamics and superfluid ring excitations in binary
Bose-Einstein condensates, Phys. Rev. Lett. \textbf{99}, 190402 (2007); R.
Zamora-Zamora, M. Lozada-Hidalgo, S. F. Caballero-Ben\'{\i}tez, and V.
Romero-Roch\'{\i}n, Vortices on demand in multicomponent Bose-Einstein
condensates, Phys. Rev. A \textbf{86}, 053624 2012).

\bibitem{arxiv} R. Driben, V. V. Konotop, and T. Meier, Precession and
nutation dynamics of nonlinearly coupled non-coaxial three-dimensional
matter wave vortices, Sci. Rep. \textbf{6}, 22758 (2016).

\bibitem{GA} S. Brenner and R. L. Scott, \textit{The Mathematical Theory of
Finite Element Methods}, Springer-Verlag (New York, 2002); J. L. Guermond,
P. Minev, and J. Shen, An overview of projection methods for incompressible
flows, Comp. Methods Appl. Mech. Engineering \textbf{195}, 6011-6045 (2006).

\bibitem{Feshbach+rf} A. M. Kaufman, R. P. Anderson, T. M. Hanna, E.
Tiesinga, P. S. Julienne, and D. S. Hall, Radio-frequency dressing of
multiple Feshbach resonance, Phys. Rev. A \textbf{80}, 050701(R) (2009).

\bibitem{Feshbach2} S. Tojo, Y. Taguchi, Y. Masuyama, T. Hayashi, H. Saito,
and T. Hirano, Controlling phase separation of binary Bose-Einstein
condensates via mixed-spin-channel Feshbach resonance, Phys. Rev. A \textbf{%
82}, 033609 (2010).

\bibitem{Shlyap} D. J. Papoular, G. V. Shlyapnikov, and J. Dalibard,
Microwave-induced Fano-Feshbach resonances, Phys. Rev. A \textbf{81},
041603(R) (2010); T. V. Tscherbul, T. Calarco, I. Lesanovsky, R. V. Krems,
A. Dalgarno, and J. Schmiedmayer, rf-field-induced Feshbach resonances,
Phys. Rev. A \textbf{81}, 050701(R) (2010).

\bibitem{Australia} M. Egorov, B. Opanchuk, P. Drummond, B. V. Hall, P.
Hannaford, and A. I. Sidorov, Measurement of \textit{s}-wave scattering
lengths in a two-component Bose-Einstein condensate, Phys. Rev. A \textbf{87}%
, 053614 (2013).


\bibitem{Tosi} B. D. Esry, C. H. Greene, J. P. Burke, Jr., and J. L. Bohn,
Hartree-Fock theory for double condensates, Phys. Rev. Lett. \textbf{78},
3594-3597 (1997); M. L. Chiofalo, S. Succi, and M. P. Tosi, Ground state of
trapped interacting Bose-Einstein condensates by an explicit imaginary-time
algorithm. Phys. Rev. E \textbf{62}, 7438-7444 (2000); W. Bao and Q. Du,
Computing the ground state solution of Bose-Einstein condensates by a
normalized gradient flow, SIAM\ J. Sci. Comput. \textbf{25}, 1674-1697
(2004).

\bibitem{spinor} T. Isoshima, M. Nakahara, T. Ohmi, and K. Machida, Creation
of a persistent current and vortex in a Bose-Einstein condensate of
alkali-metal atoms, Phys. Rev. A \textbf{61}, 063610 (2000); Y. Kawaguchi
and M. Ueda, Spinor Bose-Einstein condensates, Phys. Rep. \textbf{520},
253-381 (2012).
\end{thebibliography}
\end{document}